\documentclass[preprint,showpacs,preprintnumbers,amsmath,amssymb]{revtex4}
\usepackage{amsmath}
\usepackage{graphicx}
\usepackage{dcolumn}
\usepackage{bm}
\usepackage{tikz}
\usepackage{pgflibraryarrows}
\usepackage{pgflibrarysnakes}
\usepackage{graphicx}
\usepackage{subfigure}

\begin{document}

\providecommand{\keywords}[1]              %
{                                          %
  \small	                               %
  \textbf{\textit{Keywords---}} #1         %
}                                          %
                                           %
{\bf Submitted to Chinese Physics C.}

\title{Pseudo-rapidity distribution from a perturbative solution of viscous hydrodynamics for heavy ion collisions at RHIC and LHC}

\author{Ze Fang Jiang $^{~1,2}$}
\email{jiangzf@mails.ccnu.edu.cn}
\author{C.~B.~Yang$^{~1,2}$}
\email{cbyang@mail.ccnu.edu.cn}
\author{Chun-Bin Yang$^{~1,2}$}
\author{Chi Ding$^{~1,2}$}
\author{Xiang-Yu Wu$^{~1,2}$}

\affiliation{$^1$ Key Laboratory of Quark and Lepton Physics, Ministry of Education, Wuhan, 430079, China}
\affiliation{$^2$ Institute of Particle Physics, Central China Normal University, Wuhan 430079, China}

\begin{abstract}
The charged-particle's final state spectrum is derived from an analytic perturbative solution for the relativistic viscous hydrodynamics. By taking into account the longitudinal acceleration effect in relativistic viscous hydrodynamics, the pseudorapidity spectrum describes well the nucleus-nucleus colliding systems at RHIC and LHC. Based on both the extracted longitudinal acceleration parameters $\lambda^{*}$ and a phenomenological description of the $\lambda^{*}$, the charged-particle's pseudorapidity distributions for $\sqrt{s_{NN}}$ = 5.44 TeV Xe+Xe collisions are computed from the final state expression in a limited space-time rapidity $\eta_{s}$ region.
\end{abstract}

\keywords{ viscous hydrodynamics; final state observable; pseudorapidity distributions; longitudinal acceleration effect; xeon+xeon collisions.}

\pacs{20. 24, 20.25, 25.75}
\maketitle
\date{\today}

\section{Introduction}
Relativistic hydrodynamics seems to be an efficient tool to study
the expansion and many non-equilibrium properties of quark-gluon plasma (QGP) produced
in high energy heavy ion collisions, such as those at the Relativistic Heavy Ion Collider (RHIC) at
Brookhaven National Laboratory, USA and the Large Hadron Collider (LHC) at CERN \cite{Bass1998vz,Gyulassy:2004zy,Heinz2013th}.

There has been tremendous theoretical~\cite{Romatschke:2017ejr,Israel:1979wp,AM:2004prc,Baier:2007ix,Bhattacharyya:2008jc,Koide:2006ef,PeraltaRamos:2009kg,
Denicol:2012cn,Landau:1953gs,Hwa:1974gn,Bjorken:1982qr,Biro:2000nj,Csorgo:2003rt,Csorgo:2006ax,Csorgo:2008prc,
Nagy:2009eq,Gubser:2010ze,Gubser:2010ui,Jiang:2014uya,Jiang:2014wza,Hatta:2014gqa,Hatta:2014gga,Wu:2016pmx} and numerical work
 \cite{Schenke:2010rr,Song:2010mg,Werner:2010aa,Pang:2018zzo,Chen:2017zte}
in solving relativistic hydrodynamic equations, and those works not only simulate the fluid's dynamical evolution but also play an important role in extracting the transport properties
of the strongly coupled matter. In our previous papers~\cite{Csand:2016arx,Jiang2017}, a series of exact solutions for the relativistic accelerating perfect fluid were presented and
served as a reliable reference to study the longitudinal acceleration effect, pseudorapidity distributions and the initial state properties for colliding systems at RHIC and at LHC~\cite{Biro:2000nj,Csorgo:2006ax,Nagy:2009eq,Csand:2016arx,Jiang2017}.

In this paper,
we expand the current knowledge of accelerating hydrodynamics~\cite{Csorgo:2006ax,Csand:2016arx} by including the first-order viscous (Navier-Stokes limit) corrections in the relativistic domain.
Based on a perturbative solution~\cite{Jiang2018} which includes the longitudinal acceleration in relativistic viscous hydrodynamics, the final state observations which contain the longitudinal acceleration effect are derived.
We find that the final state observations depend on the longitudinal acceleration parameter $\lambda^{*}$, and the pseudorapidity distribution can be not only used to compare with the experimental data from the $\sqrt{s_{NN}}$ =130 GeV RHIC Au+Au collisions to the $\sqrt{s_{NN}}$ =5.02 TeV LHC Pb+Pb collisions~\cite{PHOBOS:2010re,ALICE:2016dn,Adam:2016ddh
}, but also applied to academically study the longitudinal acceleration effect for nucleus-nucleus collisions at different $\sqrt{s_{NN}}$. Furthermore, motivated by the $^{129}$Xe+$^{129}$Xe run with $\sqrt{s_{NN}}$=5.44 TeV at the LHC in October 2017 and many other prospective Xe+Xe collisions' studies\cite{Giacalone:2017dud,Eskola:2017bup,Kisslinger:2018izb}, the charged-particle's pseudorapidity distribution for the most central Xe+Xe collisions with $\sqrt{s_{NN}}$=5.44 TeV is computed with this relativistic viscous hydrodynamics model and compared with the last ALICE Collaboration data~\cite {Acharya:2018hhy}.

The organization of the paper is as follows. In Sec. 2,
The transverse momentum spectrum and (pseudo-)rapidity spectrum are derived from a perturbative solution for the relativistic viscous hydrodynamics. In Sec. 3, pseudorapidity spectrum is academically used to study the longitudinal accelerating parameter $\lambda^{*}$ at RHIC and LHC. Brief summary and discussion are given in Sec. 4.

\section{Perturbative solution and final state observables}

The basic formulation of relativistic hydrodynamics can be found in the literature~\cite{AM:2004prc,Nagy:2009eq,Paul2009}.
In this paper, we consider a system wit net conservative charge ($\mu_{i}=0$).
The flow velocity field is normalized to unity, $u^{\mu}u_{\mu}=1$ and the metric tensor is chosen as $g_{\mu\nu}$=$diag(1,-1,-1,-1)$.

The energy-momentum tensor $T^{\mu\nu}$ of the fluid system in the presence of viscosity is\\
\begin{equation}
\begin{aligned}
T^{\mu\nu}=\varepsilon u^{\mu}u^{\nu}-P\Delta^{\mu\nu}+\Pi^{\mu\nu},
\label{1}
\end{aligned}
\end{equation}
where $\varepsilon$ and $P$ are the energy density and the local isotropic pressure, respectively. The equation of state (EoS) is $\varepsilon=\kappa P$. Viscous stress tensor $\Pi^{\mu\nu}=\pi^{\mu\nu}-\Delta^{\mu\nu}\Pi$, where $\Pi$ is the bulk pressure, and $\pi^{\mu\nu}$ is the stress tensor~\cite{AM:2004prc}.
 $\Delta^{\mu\nu}=g^{\mu\nu}-u^{\mu}u^{\nu}$ and $\Delta^{\mu\nu}u_{\nu}=0$.

Following the conservation equations $\partial_{\mu}T^{\mu\nu}=0$, the energy equation and Euler equation are reduced to
\begin{equation}
\begin{aligned}
D \varepsilon&=-(\varepsilon+P+\Pi)\theta+\sigma_{\mu\nu}\pi^{\mu\nu},
\label{2}
\end{aligned}
\end{equation}
\begin{equation}
\begin{aligned}
(\varepsilon+P+\Pi)Du^{\alpha}&=\nabla^{\alpha}(P+\Pi)-\Delta^{\alpha}_{\nu}u_{\mu}D\pi^{\mu\nu}-\Delta^{\alpha}_{\nu}\nabla_{\mu}\pi^{\mu\nu},
\label{3}
\end{aligned}
\end{equation}
where shorthand notations for the differential operators $D=u^{\mu}\partial_{\mu}$, $\nabla^{\alpha}=\Delta^{\mu\alpha}\partial_{\mu}$ are introduced, with the expansion rate $\theta=\nabla_{\mu}u^{\mu}$. The shear tensor is
\begin{equation}
\begin{aligned}
\sigma^{\mu\nu}&\equiv\partial^{\langle\mu}u^{\nu\rangle}\equiv(\frac{1}{2}(\Delta^{\mu}_{\alpha}\Delta^{\nu}_{\beta}+\Delta^{\mu}_{\beta}\Delta^{\nu}_{\alpha})
-\frac{1}{d}\Delta^{\mu\nu}\Delta_{\alpha\beta})\partial^{\alpha}u^{\beta}.
\label{4}
\end{aligned}
\end{equation}

Based on the Gibbs thermodynamic relation and the second law of thermodynamics,
the simplest way to satisfy the constraint (entropy must always increase locally) is to impose the linear relationships between the thermodynamic forces and fluxes (in the Navier-Stokes limit~\cite{Israel:1979wp,AM:2004prc,Hatta:2014gga,Paul2009,DTeaney,Damodaran:2017ior,Weinberg}),
\begin{equation}
\begin{aligned}
~~\Pi=-\zeta\theta,~~~\pi^{\mu\nu}=2\eta\sigma^{\mu\nu},
\label{5}
\end{aligned}
\end{equation}
where the bulk viscosity $\zeta$ and the shear viscosity $\eta$ are two positive coefficients. Note that throughout this work we denote the shear viscosity as $\eta$, the space-time rapidity as $\eta_{s}$ and the particle pseudorapidity as $\eta_{p}$.

In the accelerating frame, the proper time $\tau=\sqrt{t^{2}-r^{2}}$ and space-time rapidity $\eta_{s}=\frac{1}{2}\log((t+r)/(t-r)) $ are independent variables, where $r=\sqrt{\Sigma_i r_i^2}$ and $x^\mu=(t,r_1,\dots,r_d)$~\cite{Csorgo:2006ax,Jiang2017}.
The velocity field is parameterized as $v=\tanh(\Omega(\eta_{s}))$,  with the flow-element rapidity $\Omega(\eta_{s})$ beings an arbitrary function of the coordinate $\eta_{s}$ and $\Omega$ is independent of the proper time $\tau$ ~\cite{Csorgo:2008prc}. (For simplicity, in the following we will use $\Omega$ to represent $\Omega(\eta_{s})$, and $\Omega'$ stands for $d\Omega/d\eta_{s}$).  With the special condition $\Omega(\eta_{s})=\eta_{s}$, it can naturally come back to the Hwa-Bjorken case~\cite{Bjorken:1982qr}.

The energy equation~Eq.(\ref{2}) and the Euler equation~Eq.(\ref{3}) in Rindler coordinate reduce to the following two differential equations,\\
\begin{equation}
\begin{aligned}
&\tau\frac{\partial T}{\partial\tau}+\tanh(\Omega-\eta_{s})\frac{\partial T}{\partial\eta_{s}}+\frac{\Omega'}{\kappa}T=\frac{\Pi_{d}}{\kappa}\frac{\Omega'^{2}}{\tau}\cosh(\Omega-\eta_{s}),\\
\label{energy1}
\end{aligned}
\end{equation}
\begin{equation}
\begin{aligned}
\tanh(\Omega-\eta_{s})&\left[\tau\frac{\partial T}{\partial\tau}+T\Omega'\right]=\frac{\Pi_{d}}{\tau}(2\Omega'(\Omega'-1)-\frac{\partial T}{\partial\eta_{s}}\\
&+\Omega''\coth(\Omega-\eta_{s}))\sinh(\Omega-\eta_{s}),
\label{euler1}
\end{aligned}
\end{equation}
where $T$ is the temperature, $\varepsilon \propto T^{\kappa+1}$, $\Pi_{d}\equiv\left(\frac{\zeta}{s}+\frac{2\eta}{s}(1-\frac{1}{d})\right)$ is a combination of the shear viscosity and bulk viscosity, $s$ the entropy density, $d$ the space dimension. $\eta/s$ and $\zeta/s$ are assumed to be constant in the following calculations~\cite{Song:2010mg,Meyer:2007dy}.
If $\Pi_{d}=0$, the exact acceleration solutions of ideal hydrodynamic have already been presented in Refs.~\cite{Csorgo:2006ax,Jiang2017}.

We are interested in the viscous case ($\Pi_{d}\neq 0$) throughout this paper. One can see that it is difficult to find a general exact analytical solution for two partial differential equations Eqs.(\ref{energy1}-\ref{euler1}) for arbitrary $\Omega$. Fortunately, one can find a perturbative solution of these energy and Euler equations. Based on the results from the ideal hydro~\cite{Csand:2016arx,Jiang2017}, we assume the case in which $\Omega\equiv\lambda\eta_{s}\equiv(1+\lambda^{*})\eta_{s}$, with $\lambda^{*}$ being the very small constant acceleration parameter ($0<\lambda^{*}\ll1$). And $\Omega'$ approximately characterizes the longitudinal acceleration of flow element in the medium.

Up to the leading order ${\mathcal O}(\lambda^{*})$, the combination of energy equation Eq.(\ref{energy1}) and Euler equation Eq.(\ref{euler1}) yields a partial differential equation depending on $\tau$ only, the temperature solution $T(\tau,\eta_{s})$ is
\begin{equation}
\begin{aligned}
T(\tau,\eta_{s})&=T_{1}(\eta_{s})\left(\frac{\tau_{0}}{\tau}\right)^{\frac{1+\lambda^{*}}{\kappa}}
+\frac{(2\lambda^{*}+1)\Pi_{d}}{(\kappa-1)\tau_{0}}\left(\frac{\tau_{0}}{\tau}\right)^{\frac{1+\lambda^{*}}{\kappa}}
\left[1-\left(\frac{\tau_{0}}{\tau}\right)^{1-\frac{1+\lambda^{*}}{\kappa}}\right],
\label{tt1}
\end{aligned}
\end{equation}
where $\tau_{0}$ is the value of proper time, $T_{1}(\eta_{s})$ is an unfixed function.

Putting Eq.(\ref{tt1}) into the Euler equation Eq.(\ref{euler1}), up to $\mathcal{O}(\lambda^{*})$, one gets\\
\begin{equation}
\begin{aligned}
T_{1}(\eta_{s})&=T_{0}\exp[-\frac{1}{2}\lambda^{*}(1-\frac{1}{\kappa})\eta_{s}^{2}]
-\frac{\left(1-\exp[-\frac{1}{2}\lambda^{*}(1-\frac{1}{\kappa})\eta_{s}^{2}] \right)\Pi_{d}}{(\kappa-1)\tau_{0}},
\label{t1}
\end{aligned}
\end{equation}
where $T_{0}$ define the values for temperature at the proper time $\tau_{0}$ and coordinate rapidity $\eta_{s}=0$.

Finally, inputting Eq.(\ref{t1}) into Eq.(\ref{tt1}), we obtain a perturbative analytical solution of the $1+1$ D embed $1+3$ D accelerating relativistic viscous hydrodynamics,\\
\begin{equation}
\begin{aligned}
T(\tau,\eta_{s})&=T_{0}\left(\frac{\tau_{0}}{\tau}\right)^{\frac{1+\lambda^{*}}{\kappa}}
{\bigg [}\exp(-\frac{1}{2}\lambda^{*}(1-\frac{1}{\kappa})\eta_{s}^{2})
+\frac{R_{0}^{-1}}{\kappa-1}{\bigg (}2\lambda^{*}+\exp[-\frac{1}{2}\lambda^{*}(1-\frac{1}{\kappa})\eta_{s}^{2}]\\
&-(2\lambda^{*}+1)
\left(\frac{\tau_{0}}{\tau}\right)^{\frac{\kappa-\lambda^{*}-1}{\kappa}} {\bigg )}
 {\bigg ]},
\label{t2}
\end{aligned}
\end{equation}
where the Reyonlds number is $R^{-1}_{0}=\frac{\Pi_{d}}{T_{0}\tau_{0}}$~\cite{AM:2004prc,Kouno:1989ps}.

The profile of $T(\tau,\eta_{s})$ is a (1+1) dimensional scaling solution in (1+3)
dimensions and it contains not only acceleration but also the viscosity dependent terms now, and the $\eta_{s}$ dependence is of the Gaussian form.
Note that when $\lambda^{*}=0$ and $R_{0}^{-1}=0$, one obtains the same solutions as the ideal hydrodynamic~\cite{Bjorken:1982qr}, when $\lambda^{*}=0$ and $R_{0}^{-1}\neq0$, one obtains the first order Bjorken solutions~\cite{AM:2004prc,DTeaney}, if $\lambda^{*}\neq0$ and $R_{0}^{-1}=0$, one obtains a special solution which is consistent with the CNC solutions' case (c) in~\cite{Csorgo:2006ax,Csorgo:2008prc}.  Furthermore, this temperature profile Eq.(\ref{t2}) implies that for a non-vanishing acceleration $\lambda^{*}$, the cooling rate is larger than for the ideal case. Meanwhile, a non-zero viscosity makes the cooling rate smaller than for the ideal case~\cite{Jiang2018}.

The thermal spectrum of charged particles at proper time $\tau_{f}$ can be obtained from the Cooper-Frye flux term~\cite{Cooper1974} and above temperature profile Eq.(\ref{t2}),\\
\begin{equation}
\begin{aligned}
\frac{d^{2}N}{2\pi p_{T}dp_{T}dy}=\frac{g}{(2\pi)^{3}}\int p_{\mu}d \Sigma^{\mu} f,
\label{20}
\end{aligned}
\end{equation}
where $g$ is the spin-degeneracy factor, $\Sigma$ is the freeze-out hypersurface.

For a system out of equilibrium, the particle phase-space distribution function $f$ contains not only the equilibrium distribution function $f_{0}$ but also the viscous corrections function $\delta f$.
In a Boltzmann approximation, $f_{0}$ and $\delta f$ have been derived in Refs. \cite{Dusling2007gi,Dusling2009df},
\begin{equation}
\begin{aligned}
f_{0}={\rm exp}\left(\frac{\mu_{i}(x)}{T}-\frac{p_{\mu}u^{\mu}}{T}\right),
\label{21}
\end{aligned}
\end{equation}
\begin{equation}
\begin{aligned}
\delta f=\frac{1}{2({\color{red}\varepsilon}+P)T^{2}}f_{0}p^{\mu}p^{\nu}\left[\pi_{\mu\nu}-\frac{2}{5}\Pi\Delta_{\mu\nu}\right].
\label{22}
\end{aligned}
\end{equation}
where $\mu_{i}$ is the chemical potential and $T$ is the temperature. When net conservative charge is considered, $\mu_{i}=0$, the particle momentum $p^{\mu}$ can be written as
\begin{equation}
\begin{aligned}
p^{\mu}=(m_{T}\cosh y, p_{T}\cos\phi, p_{T}\sin\phi,m_{T}\sinh y),
\label{23}
\end{aligned}
\end{equation}
with the transverse momentum $p_{T}$, transverse mass $m_{T}=\sqrt{p _{T}^{2}+m^{2}}$, $m$ the charged particle mass, $y$ the particle rapidity, and azimuthal angle $\phi$. For an expanding Boltzmann gas, the thermal distribution is $f_{0}=\exp[-m_{T}\cosh(\Omega-y)/T]$. The freeze-out
condition is assumed that the temperature at $\eta_{s}=0$ drops below a given $T_{f}$ value, and at the mean time the four-velocity is pseudo-orthogonal to the freeze-out hypersurface~\cite{Csorgo:2006ax,Csanad2004mm,Csanad2005qr}. The freeze-out condition satisfied $\left(\frac{\tau_{f}}{\tau}\right)^{\Omega'-1}\cosh((\Omega'-1)\eta_{s})=1$, and the integration measurement is
\begin{equation}
\begin{aligned}
p_{\mu}d\Sigma^{\mu}=m_{T}\tau_{f}\cosh^{\frac{2-\Omega'}{\Omega'-1}}((\Omega'-1)\eta_{s})\cosh(\Omega-y)rdrd\phi d\eta_{s}.
\label{25}
\end{aligned}
\end{equation}

From the leading order condition Eq.(\ref{t2}), the Cooper-Frye integration gives the thermal spectrum for the equilibrium state from an expanding cylinder geometry as
\begin{equation}
\begin{aligned}
\frac{d^{2}N^{(0)}}{2\pi p_{T}dp_{T}dy}&=\frac{1}{(2\pi)^{3}}\int p_{\mu}d \Sigma^{\mu} f_{0}=\frac{1}{(2\pi)^{3}}\int_{0}^{R_{0}}rdr\int_{0}^{2\pi}d\phi\int_{-\infty}^{+\infty}d\eta_{s}\\
~~~&\times m_{T}\tau_{f}\cosh^{\frac{1-\lambda^{*}}{\lambda^{*}}}(\lambda^{*}\eta_{s})\cosh((\lambda^{*}+1)\eta_{s}-y)\\
~~~&\times\exp\left[-\frac{m_{T}}{T(\tau,\eta_{s})}\cosh((\lambda^{*}+1)\eta_{s}-y)\right],
\label{26}
\end{aligned}
\end{equation}.

In the first order (Navier-Stokes) approximation and with Gibbs relation $\varepsilon+P=Ts$, the first-order viscous correction to the spectrum is
\begin{equation}
\begin{aligned}
\frac{d^{2}N^{(1)}}{2\pi p_{T}dp_{T}dy}&=\frac{\pi R_{0}^{2}}{(2\pi)^{3}}\int_{-\infty}^{+\infty}d\eta_{s}\frac{(1+\lambda^{*})m_{T}\cosh((\lambda^{*}+1)\eta_{s}-y)}{T^{3}(\tau,\eta_{s})}\\
&\times\exp\left[-\frac{m_{T}}{T(\tau,\eta_{s})}\cosh((\lambda^{*}+1)\eta_{s}-y)\right]\\
&\times{\bigg[}\frac{1}{3}\frac{\eta}{s}(p_{T}^{2}-2m_{T}^{2}\sinh^{2}((\lambda^{*}+1)\eta_{s}-y))\\
&~~~~~+\frac{1}{5}\frac{\zeta}{s}(p_{T}^{2}+\sinh^{2}((\lambda^{*}+1)\eta_{s}-y)){\bigg]}.
\label{28}
\end{aligned}
\end{equation}
Finally, an analytical expression of the transverse momentum distribution can be written as
\begin{equation}
\begin{aligned}
\frac{d^{2}N}{p_{T}dp_{T}dy}&=\frac{\pi R_{0}^{2}}{(2\pi)^{3}}\int_{-\infty}^{+\infty}d\eta_{s}m_{T}\cosh((\lambda^{*}+1)\eta_{s}-y)\\
&\times\exp\left[-\frac{m_{T}}{T(\tau,\eta_{s})}\cosh((\lambda^{*}+1)\eta_{s}-y)\right]\\
&\times {\bigg(}\tau_{f}\cosh^{\frac{1-\lambda^{*}}{\lambda^{*}}}
(\lambda^{*}\eta_{s})+\frac{1+\lambda^{*}}{T^{3}(\tau,\eta_{s})}\\
&\times{\bigg[}\frac{1}{3}\frac{\eta}{s}(p_{T}^{2}-2m_{T}^{2}
\sinh^{2}((\lambda^{*}+1)\eta_{s}-y))\\
&~~~~~~~+\frac{1}{5}\frac{\zeta}{s}(p_{T}^{2}+m_{T}^{2}\sinh^{2}((\lambda^{*}+1)\eta_{s}-y)){\bigg]}{\bigg)},
\label{dnptdpt}
\end{aligned}
\end{equation}
where we can find that the transverse momentum distribution is sensitive to the viscosity.

One can make use of  the condition $p_{T}\approx m_{T}$ (which is a good approximation for pions) and perform the integral of transverse momentum $p_{T}$, then the rapidity distribution can be derived as
\begin{equation}
\begin{aligned}
\frac{dN}{dy}&=\frac{\pi R_{0}^{2}}{(2\pi)^{3}}\int_{0}^{+\infty}d\eta_{s}{\bigg\{}\cosh^{\frac{1-\lambda^{*}}{\lambda^{*}}}(\lambda^{*}\eta_{s})
\frac{4\tau_{f}T^{3}(\tau,\eta_{s})}{\cosh^{2}((\lambda^{*}+1)\eta_{s}-y)}+\frac{48(1+\lambda^{*})T^{2}(\tau,\eta_{s})}{\cosh^{4}((\lambda^{*}+1)\eta_{s}-y)}\\
&\times
\left[\frac{1}{3}\frac{\eta}{s}(1-2\sinh^{2}((\lambda^{*}+1)\eta_{s}-y))+\frac{1}{5}\frac{\zeta}{s}\cosh^{2}((\lambda^{*}+1)\eta_{s}-y) \right]{\bigg\}}.
\label{dndy}
\end{aligned}
\end{equation}

Thus, the pseudorapidity distribution can be obtained from eq.(\ref{dnptdpt}),

\begin{equation}
\begin{aligned}
\frac{dN}{d\eta_{p}}&=\frac{\pi R_{0}^{2}}{(2\pi)^{3}}\int_{-\infty}^{+\infty}d\eta_{s}\int_{0}^{+\infty}dp_{T}\sqrt{1-\frac{m^{2}}{m_{T}^{2}\cosh^{2}y}}
m_{T}p_{T}\cosh((\lambda^{*}+1)\eta_{s}-y)\\
&\times\exp\left[-\frac{m_{T}}{T(\tau,\eta_{s})}\cosh((\lambda^{*}+1)\eta_{s}-y)\right]{\bigg(}\tau_{f}\cosh^{\frac{1-\lambda^{*}}{\lambda^{*}}}
(\lambda^{*}\eta_{s})+\frac{1+\lambda^{*}}{T^{3}(\tau,\eta_{s})}\\
&\times{\bigg[}\frac{1}{3}\frac{\eta}{s}(p_{T}^{2}-2m_{T}^{2}
\sinh^{2}((\lambda^{*}+1)\eta_{s}-y))+\frac{1}{5}\frac{\zeta}{s}(p_{T}^{2}+m_{T}^{2}\sinh^{2}
((\lambda^{*}+1)\eta_{s}-y)){\bigg]}{\bigg)},
\label{dndeta}
\end{aligned}
\end{equation}

where $\eta_{p}$ is the pseudorapidity of the final hadron, and we have the relationship:
$y=\frac{1}{2}
\ln\frac{\sqrt{m^{2}+p_{T}^{2}\cosh^{2}\eta_{p}}
+p_{T}\sinh\eta_{p}}{\sqrt{m^{2}+p_{T}^{2}\cosh^{2}\eta_{p}}-p_{T}\sinh\eta_{p}}$.

\section{Pseudorapidity distributions results}
To go further and illustrate the effect of longitudinal acceleration effect on the observed final state spectra, an academically comparison between our hydrodynamic results with the RHIC and LHC data~\cite{PHOBOS:2010re,ALICE:2016dn,Adam:2016ddh} is shown in this section.
\begin{figure}[!htb]
\begin{center}
\includegraphics[width=0.6\textwidth]{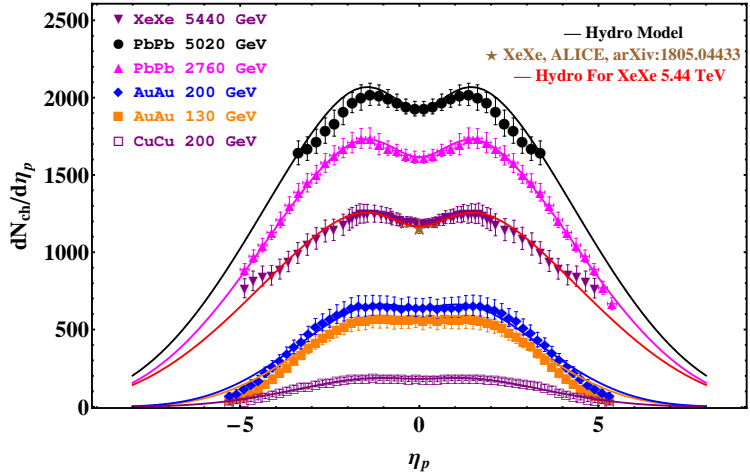}
\end{center} 
\caption{(Color online) Pseudorapidity distribution from our model calculation (the solid curves) compared to the RHIC and LHC experiment
data~\cite{PHOBOS:2010re,ALICE:2016dn,Adam:2016ddh
}. Red curve represents the pseudorapidity for Xe+Xe $\sqrt{s_{NN}}$ =5.44 TeV collision, $\lambda^{*}_{Xe+Xe}=0.030\pm0.003$ comes from Eq.(\ref{AA}), other parameters are consistent with Cu+Cu, Au+Au, and Pb+Pb. The computed central most multiplicity (five-pointed star) for Xe+Xe $\sqrt{s_{NN}}$ = 5.44 TeV is the normalization factor comes from the ALICE data~\cite{Acharya:2018hhy}, the particle's multiplicity density prediction at forward rapidity and backward rapidity come from Eq.(\ref{dndeta}).}
\label{fig1}
\end{figure}

In Fig.1, the solid curves shows the calculated pseudo-rapidity distribution, the normalization factor come from the most central multiplicity $dN/d\eta_{p}~(\eta_{p}=\eta_{0})$ with the parameters $\eta/s$=0.16~\cite{Song:2010mg}, $\zeta/s$=0.015~\cite{Meyer:2007dy}, $T_{f}=140$ MeV. $\kappa=\frac{1}{c_{s}^{2}}=\frac{\partial\varepsilon}{\partial P}$ comes from the EoS table of the Wuppertal-Budapest lattice QCD calculation~\cite
{Huovinen:2009yb}, for simplicity, $\kappa$ is assumed to be a linear relationship for different colliding systems in this study, we use the relationship from the EoS table that $\kappa\approx 7$ while $T\approx 140$ MeV \cite{Pang:2018zzo}. $m$=$220 \pm 20$ MeV is an approximate average mass of the final charged-particle ($\pi^{\pm}$, $K^{\pm}$, $p^{\pm}$) and it is calculated by a weighted average from the published experimental data~\cite{PHOBOS:2010re,ALICE:2016dn,Kisslinger:2018izb,Adare:2006ti}. The freeze-out proper time is chosen as $\tau_{f}=8$ fm for nucleus-nucleus collision. The rescatterings in the hadronic phase and the decays of hadronic resonance into stable hadrons are not included here. The acceptable integral region for each space-time rapidity is $-5.0\leq\eta_{s}\leq5.0$ (make sure the perturbative condition $\lambda^{*}\ll$1 is satisfied).

\begin{table}
\begin{center}
\begin{tabular*}{100mm}{c@{\extracolsep{\fill}}cccc}
\hline\hline
$\sqrt{s_{NN}}$~/[GeV] & $\quad$$\left.\frac{dN}{d\eta}\right\vert_{\eta=\eta_{0}}$ &~$\lambda^{*}$   &~$\chi^{2}/NDF$ \\
\hline
~$130~~$~Au+Au    &~563.9$\pm$59.5     &~0.076$\pm$0.003        &~9.41/53  \\
~$200~~$~Au+Au    &~642.6$\pm$61.0     &~0.062$\pm$0.002        &~12.23/53 \\
~$200~~$~Cu+Cu    &~179.5$\pm$17.5     &~0.060$\pm$0.003        &~2.41/53   \\
~$2760~$~Pb+Pb    &~1615$\pm$39.0      &~0.035$\pm$0.003        &~5.50/41 \\
~$5020~$~Pb+Pb    &~1929$\pm$47.0      &~0.032$\pm$0.002        &~33.0/27 \\
~$5440~$~Xe+Xe    &~~~~1167$\pm$26.0\cite{Acharya:2018hhy}      &~0.030$\pm$0.003      &~$-/-$  \\
\hline\hline
\end{tabular*}
\label{tab1}
\end{center}
\caption{Table of parameters from hydrodynamic fits in the text.
Centrality for Au+Au and Cu+Cu is $0-6\%$, centrality for Pb+Pb and Xe+Xe is $0-5\%$.
The Chi-square $\chi^{2}$ is calculated from the statistical uncertainly.}
\end{table}

We then extracted the longitudinal acceleration parameters $\lambda^{*}$ for 130 GeV Au+Au, 200 GeV Au+Au, 200 GeV Cu+Cu, 2.76 TeV Pb+Pb, and 5.02 TeV Pb+Pb the most central colliding systems without modifying any extra independent parameters. The extracted results of $\lambda^{*}$ are presented in Table I. From this table, we obtain the same conclusion as in Ref.\cite{Jiang2017}: the higher central mass energy $\sqrt{s_{NN}}$ leads to both smaller longitudinal parameter $\lambda^{*}$ and higher multiplicity density at forward and back pseudorapidity. However, because there are different set-up of the number of the free parameter between our paper and the Ref.\cite{Jiang2017}, the $\chi^{2}/NDF$ are different with the results in Ref.\cite{Jiang2017}.
A phenomenological expression for $\sqrt{s_{NN}}$ and $\lambda^{*}$ is computed based on the $\lambda^{*}$ results in Table I,\\
\begin{equation}
\begin{aligned}
\lambda^{*}=A\left(\frac{\sqrt{s_{NN}}}{\sqrt{s_{0}}}\right)^{-B},
\label{AA}
\end{aligned}
\end{equation}
where $\sqrt{s_{0}}=1$ GeV, $A=0.045$ and $B=0.23$ for nucleus-nucleus collisions\footnote{This $A$ and $B$ should depend on $N_{part}$, $N_{coll}$ and other parameters. The problem will become more clear after comparing with more RHIC and LHC data in the near future.}. Based on Eq.(\ref{AA}), the longitudinal acceleration parameter $\lambda^{*}$ for the 0-5\% centrality Xe+Xe $\sqrt{s_{NN}}$ = 5.44 TeV collisions are computed and we find $\lambda^{*}_{XeXe}$=0.030$\pm$0.003. Using the same parameters as did for the lower energy region and omitting other effects, the pseudorapidity distribution for Xe+Xe $\sqrt{s_{NN}}$ = 5.44 TeV collisions are computed and shown in Fig. 1. The red curve shows our model's calculations for Xe+Xe colliding system. We can see the hydrodynamics final state spectrum Eq.(\ref{dndeta}) can describe the experiment data well. On the basis of the predicted smaller longitudinal acceleration parameter$\lambda^{*}$ from Eq.(\ref{AA}), the Xe+Xe's multiplicity gradient at $\sqrt{s_{NN}}$ = 5.44 TeV presents a more smooth drop at forward and backward pseudorapidity than Cu+Cu collisions, Au+Au collisions and Pb+Pb collisions.

\section{Conclusions and discussion}

In conclusion, based on previous work~\cite{Csorgo:2006ax,Csand:2016arx,Jiang2017,Jiang2018}, a temperature profile is presented for accelerating relativistic viscous hydrodynamic equations by introducing the first-order correction to the conservation equations in Rindler coordinate. From such a temperature profile in 1+3-dimension space-time, one see that the fluid evolution is generally decelerated due to the viscosity $\Pi^{\mu\nu}$, meanwhile, the longitudinal accelerating effect of the flow-element compensates the decrease of temperature gradient. These two opposite behaviors may provide a new perspective for studying the medium evolution thermodynamical quantities in the viscous hydrodynamics, more detailed discussions are presented in Ref. \cite{Jiang2018}.

Furthermore, based on the temperature profile and the Cooper-Frye flux term~\cite{Cooper1974}, three final state spectra Eqs.(\ref{dnptdpt}, \ref{dndy}, \ref{dndeta}) are obtained in this paper. We find the pesudorapidity distribution describes the experimental data well for the most central colliding systems at RHIC and LHC. Also, as shown in Fig.1, the smaller longitudinal acceleration parameter $\lambda^{*}$, the flatter the pseudorapidity distributions. In addition, the introduced longitudinal acceleration factor $\lambda^{*}$ has an interesting trend is that $\lambda^{*}$ is  smaller for higher $\sqrt{s_{NN}}$. By using the measured longitudinal acceleration parameter $\lambda^{*}$ for different systems and assuming a phenomenological power function about $\lambda^{*}$, the pseudorapidity distribution for charged-particle in $\sqrt{s_{NN}}$ = 5.44 TeV Xe+Xe collisions are academically computed and described well the experimental data from ALICE Collaboration\cite{Acharya:2018hhy}.

Let's note, in this study, we focused on the final state observation's derivation and longitudinal acceleration effect's estimation, many conditions applied in this model are still far from the realistic heavy ion collisions. For a more realistic study based on or beyond this study, the following physical effects are important and should be taken into account:
(a) The equation of state (EoS) at different temperature should be closed to the results from the realistic lattice QCD calculation~\cite{Huovinen:2009yb};
(b) The viscosity should depend on $\sqrt{s_{NN}}$~\cite{Song:2010mg,Meyer:2007dy};
(c) Statistical weight for the Charged-particle's mass should be calculated from experimental data~\cite{PHOBOS:2010re,ALICE:2016dn,Kisslinger:2018izb,Adare:2006ti};
(d) The freeze-out hypersurface could be calculated by smooth method~\cite{Pang:2018zzo} or other methods;
(e) Resonance decay \cite{Alba:2017mqu} and rescatterings in the hadronic phase \cite{Song:2011hk} could be taken into account, and so on.
Those important effects and conditions should be studied in our future research.

The final state spectra obtained in this paper may not only shed new light on study the longitudinal acceleration effect of the medium's evolution, but also could serve as a test tool of hydrodynamic numerical codes~\cite{Pang:2018zzo} in the near future.\\

\section*{Acknowledgements}
We thank Xin-Nian Wang for a suggestion about xeon+xeon collisions. We thank Tam\'as Cs\"org\H{o}, M\'at\'e Csan\'ad, Long-Gang Pang, T.~S.~Bir{\'o}, Chao Wu for useful and inspiring discussions. We thank Wei Chen for a careful reading of this manuscript. This work was supported in part by the Sino-Hungarian bilateral cooperation program, under the Grand No.Te'T 12CN-1-2012-0016, by the financial supported from NNSF of China under grant No.11435004 and by the China CCNU PhD Fund 2016YBZZ100. Z-F. Jiang would like to thank L{\'e}vai P{\'e}ter  and Gergely G{\'a}bor Barnafoldi for kind hospitality during his stay at Winger RCP, Budapest, Hungary.

\end{document}